\documentclass{elsart}

\usepackage{amssymb,graphicx}
\begin{document}

\begin{frontmatter}

\title{Symplectic integrators for classical spin systems}

\author{Robin Steinigeweg \and}
\author{Heinz-J\"urgen Schmidt\corauthref{cor1}}
\corauth[cor1]{\ead{hschmidt@uos.de}}
\address{Universit\"at Osnabr\"uck, Fachbereich Physik,
Barbarastr. 7, 49069 Osnabr\"uck, Germany}

\begin{abstract}
We suggest a numerical integration procedure for solving the equations of motion of certain classical
spin systems which preserves the underlying symplectic structure of the phase space. Such symplectic
integrators have been successfully utilized for other Hamiltonian systems, e.~g.~for molecular dynamics
or non-linear wave equations. Our procedure rests on a decomposition of the spin Hamiltonian into a sum
of two completely integrable Hamiltonians and on the corresponding Lie-Trotter decomposition of the time evolution
operator. In order to make this method widely applicable we provide a large class of integrable
spin systems whose time evolution consists of a sequence of rotations about fixed axes. We test the proposed
symplectic integrator for small spin systems, including the model of a recently synthesized magnetic molecule,
and compare the results for variants of different order.
\end{abstract}

\end{frontmatter}



\section{Introduction}
\label{I} To calculate the time evolution of classical spin
systems is an important task in condensed matter physics. For
example, the cross section of neutron scattering at a spin system
is proportional to the Fourier transform of the time-depending
auto-correlation function, see \cite{VH}, which can often be
calculated in the classical limit. Completely integrable spin
systems are rare, that is, in most cases an analytical calculation of the time
evolution is not possible and one is lead to employ
numerical integration methods. Since classical spin systems are
instances of Hamiltonian systems, it is advisable to use
numerical integrators which preserve the underlying symplectic
structure of the phase space. Such ``symplectic integrators" have
been considered in the last decades \cite{HLW} and have been applied to a
variety of problems, ranging from
molecular dynamics \cite{VER} to the nonlinear Schr\"odinger equation \cite{ABL,IKS}.\\
Unfortunately, symplectic integrators for spin systems have only
rarely been considered in the literature, see
\cite{FHL,OMF1,OMF2}. The method of the independent time evolution
of sublattices, proposed in \cite{FHL,KBL,TKL}, is
volume-preserving but not symplectic, see \ref{CE} and \cite{FHL}.
Inspired by \cite{TKL}, we suggest to construct symplectic
integrators based on a splitting of the spin Hamiltonian into two
completely integrable Hamiltonians belonging to a special kind of
systems \cite{AGK,SS}. These systems are called ``${\mathcal
B}$-partitioned systems" and their time evolution can be
calculated as a sequence of rotations about fixed axes \cite{SS}.
This generalizes the St\"ormer/Verlet scheme based on separable
Hamiltonians of the form
$H=T({\bf p})+V({\bf q})$. \\
In section \ref{D} we provide the general definitions and results we need from analytical mechanics
(section \ref{DG}) and from the field of symplectic integrators based on Lie-Trotter decompositions
of the time evolution operator (section \ref{DSI}). The reader who is not familiar with the
differential geometric background may skip the technical details and only draw the moral that
a symplectic integrator approximates the exact time evolution by a sequence of calculable
time evolutions corresponding to auxiliary Hamiltonians. In section \ref{BPSS} we shortly
recapitulate the theory of ${\mathcal B}$-partitioned systems from \cite{SS}. In order to test our
suggestions we have implemented various variants of symplectic integrators and applied them to
selected small spin systems, see section \ref{R}. We report the fluctuation of the total energy
about its initial value as opposed to the constant drift for a non-symplectic Runge-Kutta method (RK4),
see section \ref{RTE}. For two integrable spin systems we compare the errors of the various
symplectic methods, including RK4, see sections \ref{RCB} and \ref{RCN}. Finally, we compare
the errors of five symplectic integrators for the integrable $N=5$ spin pyramid and fixed runtime, see
section \ref{RCR}. We close with a summary and outlook.


\section{Definitions and general results}
\label{D}
We will only formulate the pertinent definitions for symplectic integrators in the context
of spin systems.
For the general case there are excellent sources available in the literature, see e.~g.~\cite{ARN,AM}
for analytical mechanics and \cite{HLW} for symplectic integrators.
\subsection{Generalities}
\label{DG}

Classical spin configurations can be represented by
$N$-tuples of unit $3$-vectors
${\bf s}=(\vec{s}_1,\ldots,\vec{s}_N),\; |\vec{s}_\mu|^2=1$ for
$\mu=1,\ldots,N$. The compact, $2N$-dimensional manifold of all such configurations is the
{\it phase space} of the spin system
\begin{equation}\label{DG1}
{\mathcal P}={\mathcal P}_N=
\left\{
(\vec{s}_1,\ldots,\vec{s}_N)\left|\;
|\vec{s}_\mu|^2=1 \mbox{ for } \mu=1,\ldots,N
\right.\right\}
\;.
\end{equation}

A special coordinate system is given by the $2N$ local functions\\
$\varphi_\mu,\,z_\mu:{\mathcal P} \rightarrow \mathbb{R},$
implicitly defined by
\begin{equation}\label{DG2}
\vec{s}_\mu=
\left(
\begin{array}{lll}
\sqrt{1-z_\mu^2} \cos\varphi_\mu \\
\sqrt{1-z_\mu^2} \sin\varphi_\mu \\
z_\mu
\end{array}
\right)
,\;\mu=1,\ldots,N\;.
\end{equation}

A {\it tangent vector} of ${\mathcal P}$ at a point ${\bf s}\in{\mathcal P}$ can be represented
by an $N$-tuple ${\bf t}=(\vec{t}_1,\ldots,\vec{t}_N)$ of $3$-vectors satisfying the constraint
\begin{equation}\label{DG3}
\vec{t}_\mu\cdot\vec{s}_\mu=0,\;\mu=1,\ldots,N
\;.
\end{equation}
If ${\bf a,b}$ are two tangent vectors at ${\bf s}\in{\mathcal P}$, the assignment
\begin{equation}\label{DG4}
\omega({\bf a,b})=\sum_{\mu=1}^N (\vec{a}_\mu\times \vec{b}_\mu)\cdot\vec{s}_\mu
\end{equation}
defines a non-degenerate, closed $2$-form, that is, a {\it symplectic form} $\omega$.
In the coordinate system (\ref{DG2}) $\omega$ can locally be written in the form
\begin{equation}\label{DG5}
\omega=\sum_{\mu=1}^N
d\varphi_\mu \wedge dz_\mu
\;,
\end{equation}
hence $(\varphi_\mu,z_\mu)_{\mu=1,\ldots,N}$ are canonical coordinates w.~r.~t.~$\omega$.
The {\it volume form} $d{\mathcal P}$ is defined by
\begin{equation}\label{DG6}
d{\mathcal P}=\omega^N= \omega\wedge\omega\wedge\ldots\wedge\omega
\end{equation}
and has the local coordinate representation
\begin{equation}\label{DG7}
d{\mathcal P}=d\varphi_1\wedge dz_1\wedge\ldots\wedge d\varphi_N\wedge dz_N
\;.
\end{equation}

A smooth Hamiltonian $H:{\mathcal P}\rightarrow\mathbb{R}$ generates the
{\it Hamiltonian vector field} $X_H$ implicitly defined by
\begin{equation}\label{DG8}
i_{X_H} \;\omega \equiv\omega(X_H,\;)=dH
\;.
\end{equation}
The corresponding Hamiltonian equations of motion are
\begin{equation}\label{DG9}
\frac{d}{dt}\;{\bf s}(t) = X_H({\bf s}(t))
\end{equation}
and assume their usual form
\begin{equation}\label{DG10}
\frac{d}{dt}\varphi_\mu(t)=\frac{\partial H}{\partial z_\mu}\;,\quad\quad
\frac{d}{dt}z_\mu(t)=-\frac{\partial H}{\partial \varphi_\mu}\;,\quad \mu=1,\ldots,N\;,
\end{equation}
in the canonical coordinate system (\ref{DG2}).
By writing the solution ${\bf s}(t)$ of (\ref{DG10})
in the form ${\bf s}(t)={\mathcal F}_t(H)({\bf s}(0))$
we obtain the {\it Hamiltonian flow} ${\mathcal F}_t(H):{\mathcal P}\rightarrow {\mathcal P}$. It is defined for all
initial values ${\bf s}(0)$ and for all
$t\in\mathbb{R}$ since ${\mathcal P}$ is compact, i.~e.~$X_H$ is a {\it complete} vector field.
Analogously, the flow of a general vector field can be defined.\\

A smooth map $\phi:\;{\mathcal P}\rightarrow{\mathcal P}$ is called {\it symplectic} iff it preserves the symplectic
form, i.~e.~iff $\phi^\ast\omega=\omega$. Every symplectic map preserves the phase space volume, but not
conversely, see the counter-example below. Any Hamiltonian flow ${\mathcal F}_t(H)$ is symplectic, cf.~, for example,
theorem $8.1.9$ in \cite{AMR}. Conversely, if the flow ${\mathcal F}_t$ of a complete vector field $X$ is
symplectic, then
\begin{equation}\label{DG11}
{\mathcal L}_X \omega({\bf s})=\frac{d}{dt}
\left.\left( {\mathcal F}_t^\ast \omega \right)({\bf s})\right|_{t=0}=0
\;,
\end{equation}
where ${\mathcal L}_X$ is the Lie derivative. Hence $0={\mathcal L}_X \omega=i_X\;d\omega+di_X\omega=0+di_X\omega\;,$
that is, $i_X\omega$ is a closed $1$-form, and has, by the Poincar\'{e} lemma, locally the form
$i_X\omega=dK$. To summarize: symplectic flows are, at least locally, generated by suitable Hamiltonians $K$.

\subsection{Symplectic integrators}
\label{DSI}

\begin{table}\label{Table1}
\caption{Various decompositions of the form (\ref{DSI5}) which give rise to different symplectic integrators.
}
\begin{tabular}{|l|l|l|l|l|}
\hline
Name & Abbr. & Order &Coefficients & Ref.\\
\hline\hline
Suzuki-Trotter & ST1 & $1$ & $a_1=b_1=\frac{1}{2}$ & \cite{TR}\\
\hline
Suzuki-Trotter & ST2 & $2$ & $a_1=a_2=\frac{1}{2},\; b_1=1, b_2=0$ & \cite{STR}\\
\hline
Suzuki-Trotter & ST4 & $4$ & $a_1=a_6=\frac{p}{2},\;b_6=0 $& \cite{SU90}\\
 & & & $b_1=a_2=b_2=b_4=a_5=b_5=p $ & \\
 & & & $a_3=a_4=\frac{1-3p}{2},\; b_3=1-4p$ & \\
& & & $p=\frac{1}{4-4^{1/3}}$ & \\
\hline
Forest-Ruth & FR & $4$ & $a_1=a_4=\frac{\theta}{2},\; a_2=a_3=\frac{1-\theta}{2}$ & \cite{FR}\\
& & & $b_1=b_3=\theta,\; b_4=0$ & \\
& & & $b_2=1-2\theta,\; \theta=\frac{1}{2-2^{1/3}}$ & \\
\hline
Optimized & OFR & $4$ & $a_1=a_5=\xi,\; a_2=a_4=\chi,\;b_5=0$ & \cite{OMF}\\
Forest-Ruth & & & $a_3=1-2(\xi+\chi),\; b_1=b_4=\frac{1-2\lambda}{2}$ & \\
 &&& $b_2=b_3=\lambda=-0.09156203$ &\\
 &&& $\xi=0.17208656,\; \chi=-0.16162176$ &\\
\hline
\end{tabular}
\end{table}

From an abstract point of view, a {\it symplectic integrator} is an approximation
of some exact flow ${\mathcal F}_t(H)$ by the composition of symplectic maps
$\phi_\nu:{\mathcal P}\rightarrow {\mathcal P}$, which can be calculated analytically or numerically exact.
In this article we assume that the Hamiltonian $H$ is decomposable into completely integrable
Hamiltonians $H_i$ in the form
\begin{equation}\label{DSI1}
H=\sum_i H_i
\end{equation}
and that the $\phi_\nu$ are the Hamiltonian flows
corresponding to certain $H_i$. The precise form of the
correspondence is given by a Lie-Trotter decomposition of the flow
${\mathcal F}_t(H)$ written as an exponential operator
\begin{equation}\label{DSI2}
{\mathcal F}_t(H)=e^{t \mathcal H}
\;.
\end{equation}
In order to make sense of (\ref{DSI2})
we have to linearize the Hamiltonian equations of motion.
To this end we consider ${\mathcal F}_t(H)$
acting on functions $f:{\mathcal P}\rightarrow \mathbb{C}$ via
\begin{equation}\label{DSI3}
{\mathcal F}_t(H)^\ast f({\bf s})\equiv
f
\Big(
{\mathcal F}_{-t}(H)({\bf s})
\Big)\;.
\end{equation}
If $f$ runs through ${\mathcal L}^2({\mathcal P},d{\mathcal P})$,
the Hilbert space of (equivalence classes of) square-integrable complex functions,
(\ref{DSI3}) defines a continuous, unitary $1$-parameter group, see section $7.4$ of \cite{AMR}
for details. By Stone's theorem, this group has the form (\ref{DSI2}) with an anti-selfadjoint
operator ${\mathcal H}$. One can show that ${\mathcal H}$ can be expressed by means of the Poisson bracket
according to
\begin{equation}\label{DSI4}
{\mathcal H} f =\{ H,\;f\}\equiv\omega(X_H,X_f),\quad\quad f\mbox{ smooth}
\;,
\end{equation}
but we will not need this in the sequel.  (\ref{DSI2})
is only needed to provide a basis for using the techniques of Lie-Trotter
decomposition for Hamiltonian flows.\\

For sake of simplicity let us consider the special case $H=H_1+H_2$ and hence ${\mathcal
H}={\mathcal H}_1+{\mathcal H}_2$. We are looking for $\ell$-th order Lie-Trotter
decompositions which have the form
\begin{equation}\label{DSI5}
e^{t({\mathcal H}_1+ {\mathcal H}_2)} =
\prod_{i=1}^k e^{a_i t{\mathcal H}_1}e^{b_i t{\mathcal H}_2} + {\mathcal O}(t^{\ell+1})
\;.
\end{equation}
Both sides of (\ref{DSI5}) are expanded into power series in terms of
$t$ and set equal up to terms including $t^\ell$. This yields a
system of, in general, non-linear equations for the unknown
coefficients $a_i,\; b_i$. Except for $\ell=1$ the corresponding
solutions are not unique. Hence there exist several decompositions
and thus several symplectic integrators of the same order $\ell$.
In this article we will use the decompositions enumerated in table
1. All corresponding integrators are {\it symmetric}, or time-reversible, see \cite{HLW}.
Obviously, the Lie-Trotter decomposition (\ref{DSI5}) is a
good approximation only for small $t$. Therefore the given time
interval $[0,t]$ is usually split into $L$ intervals of length
$\Delta$ and (\ref{DSI5}) is separately applied to each time step
$\Delta$. Hence, apart from the choice of the
decomposition, $\Delta$ is a further parameter of the integration
procedure, see section \ref{R}.

\subsubsection{A counter-example}
\label{CE} It seems plausible that an arbitrary splitting
$X_H=X_1+X_2$ of a Hamiltonian vector field need not correspond to
a splitting of the Hamiltonian $H=H_1+H_2$, such that
$X_i=X_{H_i}$ for $i=1,2$. Hence the decomposition $X_H=X_1+X_2$
does not necessarily lead to symplectic integrators. Nevertheless,
we will illustrate this by an example which is connected with a
numerical integrator used for bi-partite spin systems, see
\cite{FHL,KBL,TKL}. Such spin systems can be divided into two
disjoint subsets of spins $A$ and $B$, such that the interaction
is only non-zero between spins of different subsets. The first
step of the numerical procedure consists of fixing the $A$-spins
and calculating the time evolution of all $B$-spins. In the second
step the role of $A$ and $B$ is interchanged, and so on. In a
single step each spin of one subset rotates about the fixed
(weighted) sum of all its neighboring spins; hence the numerical
integrator preserves the volume of the total phase space.
But, as we will show, this integrator is not symplectic,
see also the corresponding remark in \cite{FHL}.\\
It suffices to consider just two spins and a single step of the
described numerical integrator which solves the equations of
motion
\begin{equation}\label{CE1}
\frac{d}{dt}\vec{s}_1=\vec{s}_2\times\vec{s}_1,\quad \frac{d}{dt}\vec{s}_2=\vec{0}
\;,
\end{equation}
defining a vector field $X$ on ${\mathcal P}_2$.
We adopt canonical coordinates $\varphi_1,z_1,\varphi_2,z_2$ defined in (\ref{DG2}) and use the local
expression $\omega=d\varphi_1\wedge dz_1+d\varphi_2\wedge dz_2$ of the symplectic form. After some
elementary calculations we obtain
\begin{eqnarray}\label{CE2}
i_X \omega &=& \dot{\varphi}_1 dz_1 -\dot{z}_1 d\varphi_1\\ \nonumber
&=& \left(
z_2-z_1{\textstyle \sqrt{\frac{1-z_2^2}{1-z_1^2}}}\cos(\varphi_1-\varphi_2)
\right) dz_1\\
&&-\sqrt{(1-z_1^2)(1-z_2^2)}\sin(\varphi_1-\varphi_2) d\varphi_1
\;.
\end{eqnarray}
Obviously, $i_X \omega$ is not closed, and hence $X$ does not generate a symplectic flow,
cf.~the discussion after (\ref{DG11}).

\section{{$\mathcal B$}-partitioned spin systems}
\label{BPSS}
The symplectic integrators considered in section \ref{DSI} are based on a splitting of the
spin Hamiltonian into a sum of completely integrable Hamiltonians: $H=\sum_i H_i$.
For Heisenberg Hamiltonians
\begin{equation}\label{BPSS1}
H({\bf s})=\sum_{\mu<\nu} J_{\mu\nu} \vec{s}_\mu\cdot\vec{s}_\nu
\;,
\quad\quad
\mbox{where }
J_{\mu\nu}\in\mathbb{R}
\;,
\end{equation}
such a splitting is always possible; in fact, each summand in (\ref{BPSS1})
is a completely integrable dimer Hamiltonian.
However, it seems favorable to work with as few summands as possible,
or, equivalently, to work with ``large" integrable Hamiltonians.
To this end we will define a special class of completely integrable spin systems called
${\mathcal B}$-partitioned systems, following \cite{SS}.\\
As an example, consider the Heisenberg Hamiltonian of the spin square
\begin{equation}\label{BPSS2}
H_\square=\vec{s}_1\cdot\vec{s}_2+
\vec{s}_2\cdot\vec{s}_3+\vec{s}_3\cdot\vec{s}_4+\vec{s}_4\cdot\vec{s}_1
\;.
\end{equation}
It is integrable because it can be written as
\begin{equation}\label{BPSS3}
H_\square={\textstyle \frac{1}{2}}
\left(
(\vec{s}_1+\vec{s}_2+\vec{s}_3+\vec{s}_4)^2-(\vec{s}_1+\vec{s}_3)^2-(\vec{s}_2+\vec{s}_4)^2
\right)
\;.
\end{equation}
The grouping of the spins in (\ref{BPSS3}) can be encoded in a ``partition tree"
\begin{equation}\label{BPSS4}
{\mathcal B}_\square=\{\{1,2,3,4\},\{1,3 \},\{2,4 \},\{1 \},\{2 \},\{3 \}\,\{4 \}\}
\;.
\end{equation}
Generalizing this example, we define
\begin{defn}\label{D1}
A \underline{partition tree} ${\mathcal B}$ over a finite set $\{1,\ldots,N\}$ is a set of subsets of $\{1,\ldots,N\}$
satisfying
\begin{enumerate}
\item $\emptyset \notin {\mathcal B}$ and $\{1,\ldots,N\}\in {\mathcal B}$,
\item for all $M, M'\in {\mathcal B}$ either $M\cap M'=\emptyset$ or $M\subset M'$ or $M'\subset M$,
\item for all $M\in {\mathcal B}$ with $|M|>1$ there exist $M_1, M_2\in {\mathcal B}$ such that
$M=M_1\dot{\cup}M_2$.
\end{enumerate}
\end{defn}
It follows from definition \ref{D1} (2) that the subsets $M_1,M_2$ satisfying $M=M_1\dot{\cup}M_2$
in definition \ref{D1} (3) are unique, up to their order.
$M_1,M_2$ are hence defined for all $M\in{\mathcal B}$ with $|M|>1$.
$M_1$ and $M_2$ denote the two uniquely determined ``branches" starting from $M$.
It follows that ${\mathcal B}$ is a binary tree with the root $\{1,\ldots,N\}$ and singletons $\{\mu\}$ as leaves.
More general partitions into $k$ disjoint subsets can be reduced to subsequent binary partitions
and hence need not be considered.
For all $M\in {\mathcal B}$ there is a unique path
\begin{equation}\label{BPSS5}
{\mathcal P}_M({\mathcal B})\equiv \{M'\in{\mathcal B}\;|\;M\subset M'\}
\end{equation}
joining $M$ with the root of ${\mathcal B}$. It is linearly ordered since $M\subset M'$
and $M\subset M''$ imply $M'\subset M''$ or $M''\subset M'$ by definition \ref{D1} (2).
Especially, every element $\mu\in\{1,\ldots,N\}$ belongs to a unique, linearly ordered
{\it construction path}
\begin{equation}\label{BPSS6}
{\mathcal P}_\mu({\mathcal B})\equiv \{M\in{\mathcal B}\;|\;\mu\in M\}
\;.
\end{equation}
For $\mu\neq\nu\in\{1,\ldots,N\}$ let $M_{\mu\nu}\in{\mathcal B}$ denote the smallest set of ${\mathcal B}$ such that
$\mu,\nu\in M_{\mu\nu}$, i.~e.~$M_{\mu\nu}\in{\mathcal B}$ is
the set where both construction paths of $\mu$ and $\nu$ meet the first time.
For $M\neq\{1,\ldots,N\}$ we will denote by $\overline{M}$ the ``successor" of $M$, that is, the smallest
element of ${\mathcal P}_M({\mathcal B})$ except $M$ itself.\\

Consider real functions $J$ defined on a partition tree
\begin{equation}\label{BPSS7}
J:{\mathcal B}\longrightarrow \mathbb{R}
\end{equation}
satisfying $J(\{\mu\})=0$ for all $\mu=1,\ldots,N$.
Then
\begin{equation}\label{BPSS8}
H=\sum_{\mu<\nu}J(M_{\mu\nu}) \vec{s}_\mu\cdot \vec{s}_\nu
\end{equation}
defines a Heisenberg Hamiltonian.
The corresponding spin system will be called a {${\mathcal B}$-\it partitioned system}
or sometimes, more precisely, a
{$({\mathcal B},J)-$}{\it system}.
For example, the spin square (\ref{BPSS2}) is obtained by the partition tree
(\ref{BPSS4})
and by the function $J$ with $J(\{1,2,3,4\})=1$ and $J(M)=0$ else.\\

Let $\vec{S}_M$ denote the total spin vector of the subsystem $M\subset\{1,\ldots,N\}$
with length $S_M$. Further, let
$\mathbb{D}(\vec{\omega}, t)$  denote the $3$-dimensional rotation matrix with axis $\vec{\omega}$ and angle
$|\vec{\omega}| \, t$. In the special case $\vec{\omega}=\vec{0}$, $\mathbb{D}(\vec{\omega},t)$ denotes
the identity matrix $\mathbb{I}$.
Then the following can be proven, see \cite{SS}:
\begin{thm}\label{T1}
Let $H$ be the Hamiltonian of a $({\mathcal B},J)$-system. Then its time evolution is given by
\begin{equation}\label{BPSS9}
\vec{s}_\mu(t)=\stackrel{\leftarrow}{\prod}_{M\in{\mathcal P}_\mu({\mathcal B})}
\mathbb{D}\left(
\vec{S}_M(0), (J(M)-J(\overline{M}))t
\right)
\quad\vec{s}_\mu(0)
\;,
\;\; \mu=1,\ldots,N,
\end{equation}
where the arrow above the product symbol denotes a product according to a decreasing sequence of sets
$M\in{\mathcal P}_\mu({\mathcal B})$
from left to right and $J(\overline{M})\equiv 0$ for $M={\{1,\ldots,N\}}$.
\end{thm}
We note that the time evolution in the presence of a Zeeman term in a Hamiltonian of the form
$H+\vec{B}\cdot\vec{S}$, where $\vec{B}$ is the dimensionless magnetic field, is obtained by multiplying
(\ref{BPSS9}) from the left with $\mathbb{D}(\vec{B},t)$.\\

If we stick to the symplectic integrators of table 1 and to ${\mathcal B}$-partitioned systems as completely
integrable spin systems,  our method only applies to those spin systems whose Hamiltonian can be written
as the sum of two Hamiltonians of ${\mathcal B}$-partitioned subsystems. The spin cube with one additional
space diagonal is an example which can only be decomposed into at least three ${\mathcal B}$-partitioned subsystems.
But our method can, in principle, be extended to decompositions of the Hamiltonian into more than two summands.
As an non-trivial example where our method works without modification we mention the spin system of $N=30$ spins
which are uniformly coupled according to the edges of an icosidodecahedron, see \cite{WEB}.
Such a spin system has been physically
realized as an organic molecule containing $30$ paramagnetic Fe-ions, see \cite{FE30}.
In figure 1 the planar graph of the icosidodecahedron is decomposed
into two ${\mathcal B}$-partitioned subsystems $A$ and $B$.
$A$ consists of $6$ disjoint ``bow ties" of the form $\Join$ and
$B$ of $8$ disjoint triangles together with $6$ single spins.

\begin{figure}
  \includegraphics[width=\columnwidth]{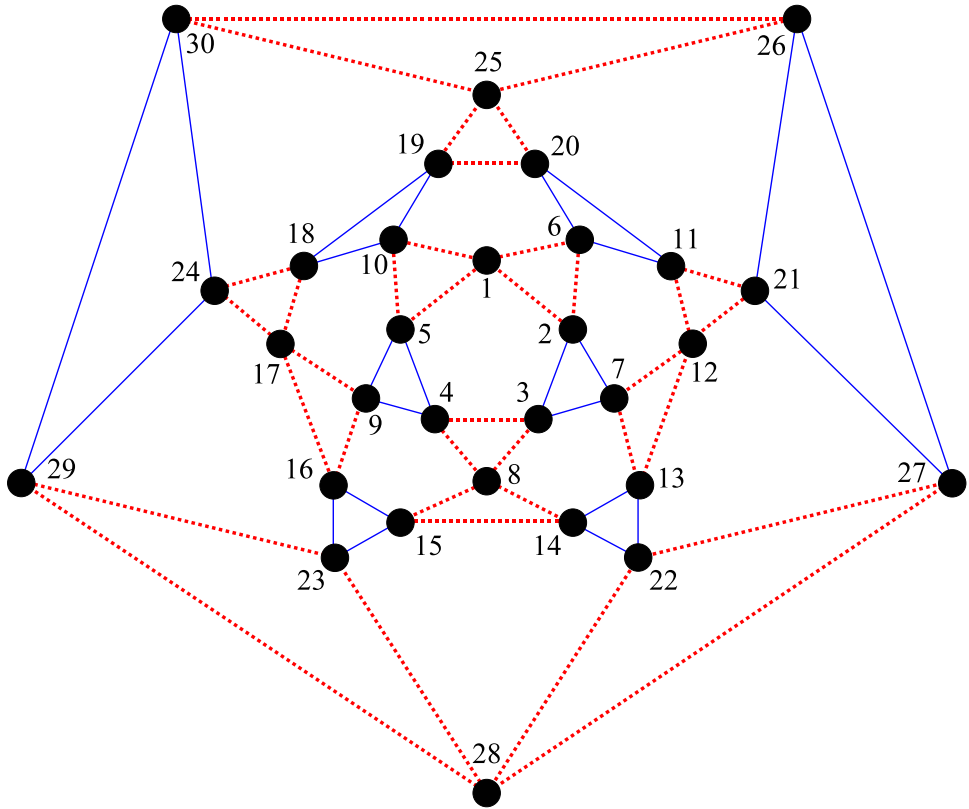}
\caption{\label{fig01}Decomposition of the graph of the icosidodecahedron into $6$ bow ties
(dashed lines) and $8$ triangles (solid lines). This is the basis of the
symplectic integrators approximately solving the equations of motion for the
corresponding classical spin system.
}
\end{figure}


\section{Results}
\label{R}
We have implemented the various symplectic integrators described above using the computer algebra software
MATHEMATICA 4.0 and have applied them to small spin systems. This seems to be sufficient in order to test general properties
of the algorithms and to compare the different decompositions according to table 1. For more extensive tests and
``real life" applications an implementation using other computer languages would be advisable.\\
For a non-integrable spin system it is impossible to compare the results of a numerical integration with the exact
result since the exact result is not known by definition. Possible tests are observations of conserved quantities
as the total energy $H$ for non-integrable spin systems or observations of non-conserved quantities for
integrable spin systems. These tests will be reported and discussed in the next subsections.

\subsection{Total energy}
\label{RTE}
Figure 2 and 3 show results of numerical integrations of the Hamiltonian equations of motion for the spin system
corresponding to the icosidodecahedron, see figure 1. We choose physical units such that the coupling constant
$J$ assumes the value $1$. For all integrations the time interval is chosen as $[0,100]$
and the time step is $\Delta=0.1$. The initial spin configuration is chosen randomly. The symplectic integrators
applied to this problem are based on a decomposition of the icosidodecahedron into $6$ bow ties
and $8$ triangles as explained above.\\
Figure 2 shows the total energy and the three components of the total spin $\vec{S}$ as a function of time
calculated by the first order integrator ST1. Whereas $\vec{S}$ is exactly conserved by all symplectic integrators
considered in this article, the total energy fluctuates about its initial value with a maximal deviation of
approximately $7\%$.\\
\begin{figure}
  \includegraphics[width=\columnwidth]{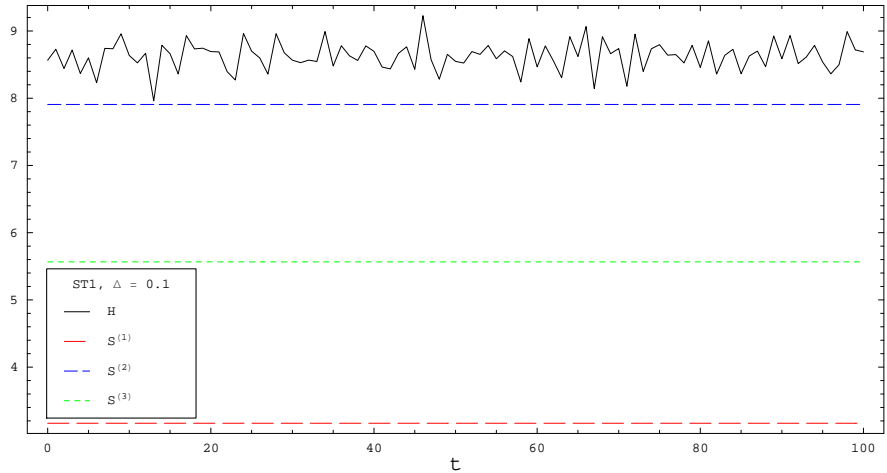}
\caption{\label{fig02}Total energy and the three components of the total spin $\vec{S}$
of the icosidodecahedron
as a function of time
calculated by ST1.
}
\end{figure}
For symplectic integrators of $4$th order the same behavior of the total energy can be observed, except that the
range of the fluctuation is much smaller. The absolute maximal deviation is about $5\cdot 10^{-4}$ for FR and
$5\cdot 10^{-5}$ for OFR and ST4, see figure 3. In contrast to these results, a $4$th order Runge-Kutta method (RK4)
yields a systematic drift of the total energy which reaches a deviation of $1.5\cdot 10^{-3}$ at $t=100$.
This is typical for non-symplectic integrators, see \cite{HLW}, and one of the main reasons to adopt symplectic
methods for Hamiltonian systems.\\

\begin{figure}
  \includegraphics[width=\columnwidth]{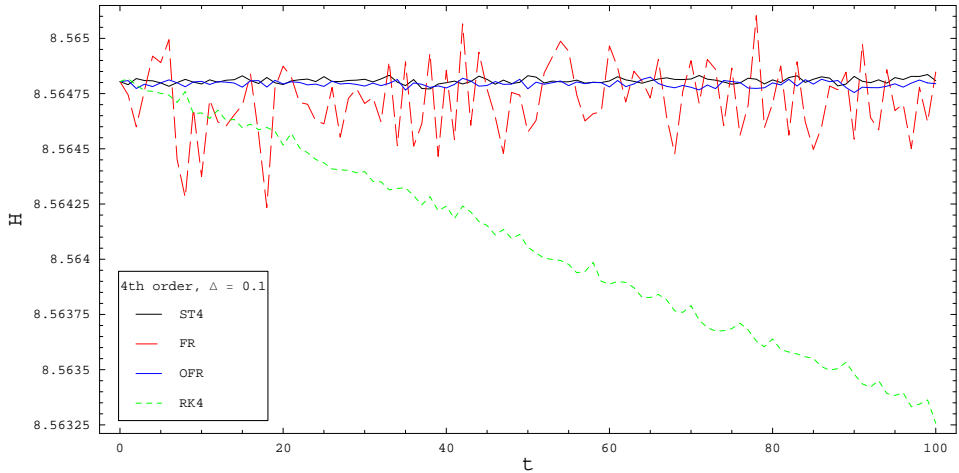}
\caption{\label{fig03}Total energy of the icosidodecahedron as a function of time
calculated by ST4, FR, OFR and RK4.
}
\end{figure}

\subsection{Comparison with exact solutions}
\label{RC}
We compare non-conserved quantities calculated by the various numerical methods with the exact solutions
for two integrable systems, the bow tie and  ``Nicholas' house", see figure 4. The latter is named
after a German nursery-rhyme (``Das ist das Haus vom Nikolaus"). The time interval $[0,100]$,
the time step $\Delta=0.1$ and the random choice of the initial configuration is similar as in the previous sections.
Although the results of the comparison with exact solutions shed some light
on the respective merits of the different methods,
it seems dangerous to generalize them to non-integrable problems where the distance between near-by solutions
may increase exponentially.

\begin{figure}
  \includegraphics[width=\columnwidth]{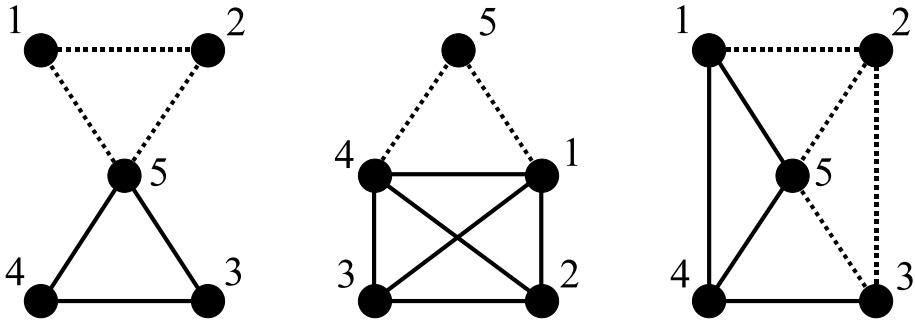}
\caption{Three integrable spin systems used for tests of numerical integrators: The bow tie,
Nicholas' house and the pyramid. The decomposition into integrable subsystems
used for symplectic integrators is indicated by solid and dashed lines.
}
\end{figure}

\subsubsection{Bow tie}
\label{RCB}
Figure 5 shows the quantity $\vec{s}_1\cdot\vec{s}_5+\vec{s}_2\cdot\vec{s}_5+\vec{s}_3\cdot\vec{s}_5$ as an exact function
of time. Figure 6 shows the absolute deviation
$\delta(\vec{s}_1\cdot\vec{s}_5+\vec{s}_2\cdot\vec{s}_5+\vec{s}_3\cdot\vec{s}_5)$
between the numerical and the exact
value in logarithmic scale for the $4$th order integrators considered above. These deviations seem to increase
linearly in time (note the logarithmic scale) but with different orders of magnitude.
The sharp minima of the logarithmic deviations in this and the following figures are due to
intersections between the exact and the approximate functions.
At $t=100$ the four integrators
can be ordered into a decreasing sequence according to their deviations, namely FR, RK4, OFR, ST4, where the
ratio between two neighbors of this sequence is approximately a factor of $10$. It is somewhat surprising
that the non-symplectic RK4 is better than FR, but w~r.~t.~conserved quantities FR should outperform RK4, as
shown in section \ref{RTE}.

\begin{figure}
  \includegraphics[width=\columnwidth]{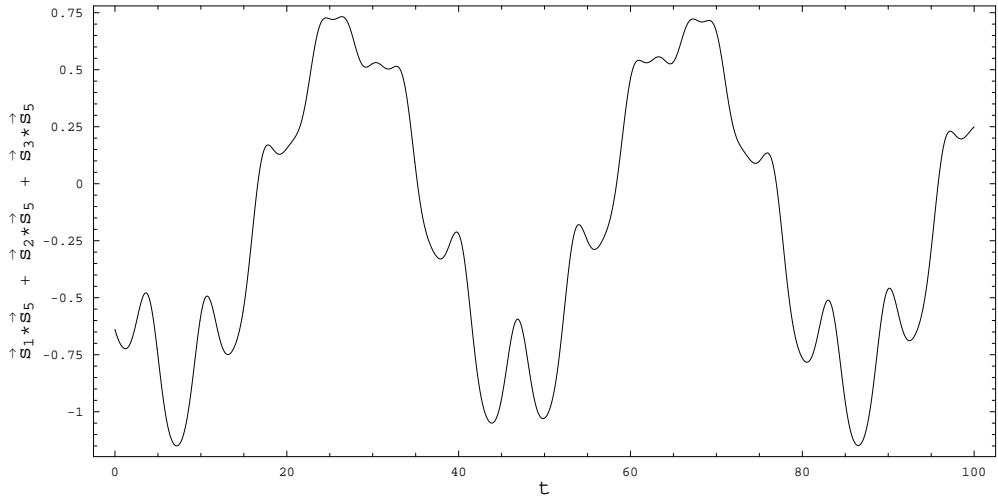}
\caption{\label{fig05}$\vec{s}_1\cdot\vec{s}_5+\vec{s}_2\cdot\vec{s}_5+\vec{s}_3\cdot\vec{s}_5$ as an exact function
of time for the bow tie.
}
\end{figure}
\begin{figure}
  \includegraphics[width=\columnwidth]{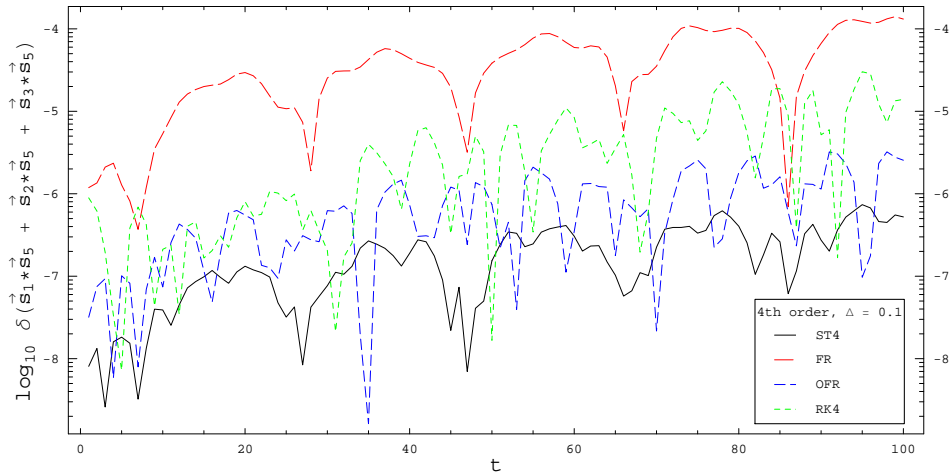}
\caption{\label{fig06}$\log_{10}\delta(\vec{s}_1\cdot\vec{s}_5+\vec{s}_2\cdot\vec{s}_5+\vec{s}_3\cdot\vec{s}_5)$
as a function of time for the bow tie and the integrators ST4, FR, OFR, and RK4.
}
\end{figure}
\subsubsection{Nicholas' house}
\label{RCN}
It is advisable to consider another example in order to see whether the above findings for the bow tie
are typical.
Figure 7 shows the quantity $\vec{s}_1\cdot\vec{s}_3+\vec{s}_2\cdot\vec{s}_4$ for the spin system called ``Nicholas' house"
as an exact function of time. Figure 8 shows the absolute deviation
$\delta(\vec{s}_1\cdot\vec{s}_3+\vec{s}_2\cdot\vec{s}_4)$
between the numerical and the exact
value in logarithmic scale for the $4$th order integrators considered above. These deviations seem to increase again
linearly in time (note the logarithmic scale). At $t=100$ we have two groups, (FR, RK4) and (OFR, ST4) with comparable
deviations within these groups, where the deviations of the second group are almost two orders of magnitude
smaller that those of the first group.

\begin{figure}
  \includegraphics[width=\columnwidth]{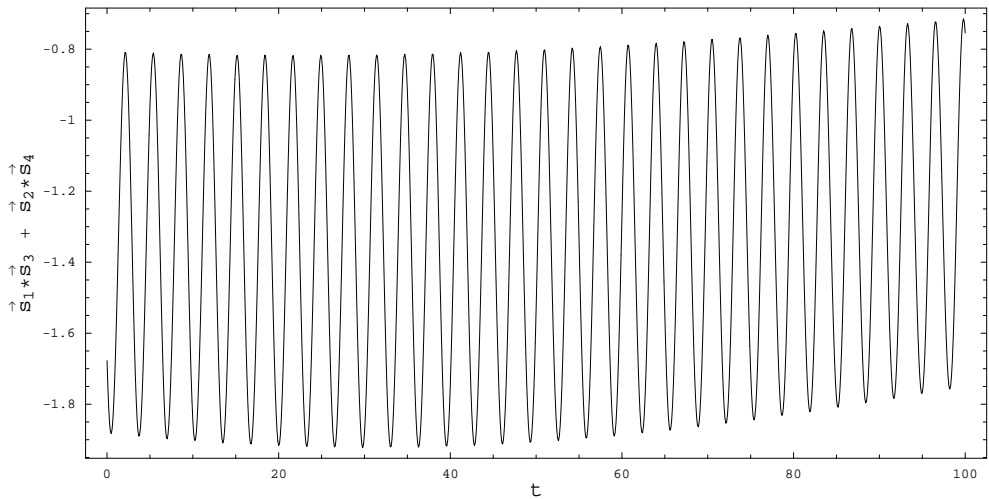}
\caption{\label{fig07}$\vec{s}_1\cdot\vec{s}_3+\vec{s}_2\cdot\vec{s}_4$ as an exact function
of time for Nicholas' house.
}
\end{figure}
\begin{figure}
  \includegraphics[width=\columnwidth]{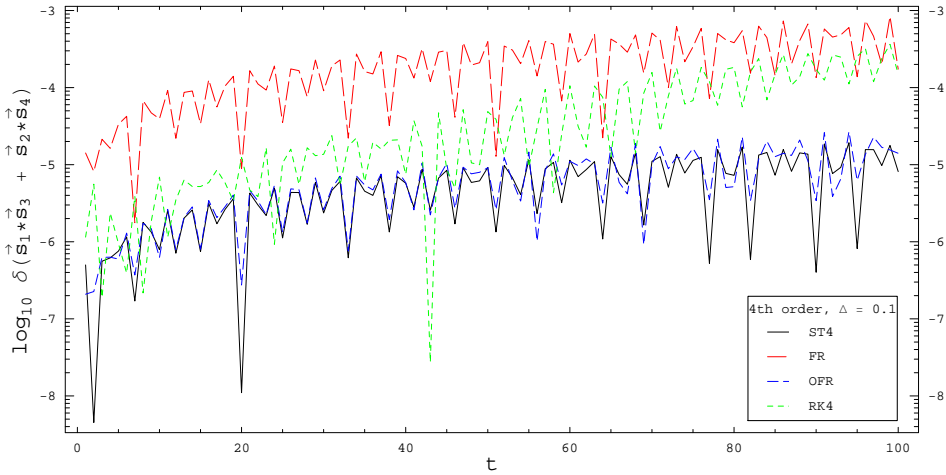}
\caption{\label{fig08}$\log_{10}\delta(\vec{s}_1\cdot\vec{s}_3+\vec{s}_2\cdot\vec{s}_4)$
as a function of time for Nicholas' house and the integrators ST4, FR, OFR, and RK4.
}
\end{figure}

\subsection{Comparison for given runtime}
\label{RCR}
From a practical point of view it is not important which numerical procedure shows the
smallest deviations for a fixed time step $\Delta$ but rather for a fixed runtime. We will provide a first
test of this kind. For this test we have to exclude the Runge-Kutta procedures since they are implemented
in the NDSolve-command of MATHEMATICA and hence their runtime cannot be compared with the symplectic integrators
programmed in MATHEMATICA code. The NDSolve-command of MATHEMATICA 5.0
also allows the choice of symplectic integrators, but these integrators
are not suited for spin systems since they rest on a splitting of the form $H=T({\bf p})+V({\bf q})$, where
${\bf p,q}$ are sets of canonical coordinates.\\

\begin{figure}
  \includegraphics[width=\columnwidth]{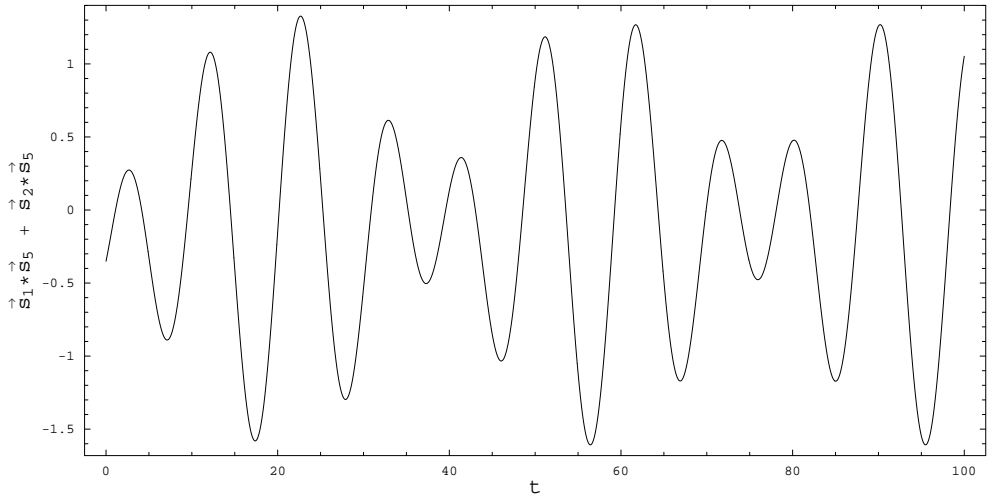}
\caption{\label{fig09}$\vec{s}_1\cdot\vec{s}_5+\vec{s}_2\cdot\vec{s}_5$ as an exact function
of time for the pyramid.
}
\end{figure}
\begin{figure}
  \includegraphics[width=\columnwidth]{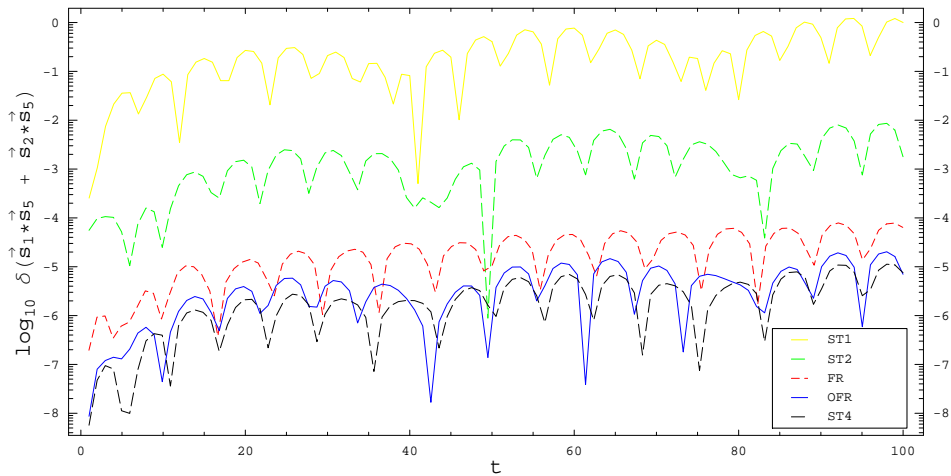}
\caption{\label{fig10}$\log_{10}\delta(\vec{s}_1\cdot\vec{s}_5+\vec{s}_2\cdot\vec{s}_5)$
as a function of time for the pyramid and the integrators ST1, ST2, ST4, FR, and OFR.
}
\end{figure}

The runtime will be measured in terms of the number of ``basic operations". A basic operation is the calculation
of the exact time evolution ${\mathcal F}_\Delta(H_i)$ for the Hamiltonian $H_i$ of an integrable subsystem. All
basic operations approximately require the same cpu-time. The common task is to calculate the quantity
$\vec{s}_1\cdot\vec{s}_5+\vec{s}_2\cdot\vec{s}_5$ as a function of $t\in [0,100]$ for a random
start configuration of the spin pyramid, see figure 4, using maximal $10.000$ basic operations.
For every numerical procedure the appropriate step size $\Delta$ is separately chosen. The results are
compared with the exact solution, see figure 9,
and the deviations are plotted as functions of $t$ in logarithmic scale, see figure 10. The deviations
vary over $8$ orders of magnitude and seem to increase with $t$. It turns out that ST4 is three orders of
magnitude more precise than ST2 and even five orders of magnitude more precise than ST1. Whereas FR lies between
ST2 and ST4, OFR is close to ST4, although its maximal deviation is about two times larger than that of ST4.
These results indicate that it might be worth while to adopt symplectic integrators of even higher order,
say, for example, ST6 or ST8.

\section{Summary and outlook}
We have proposed a symplectic integrator scheme for classical spin systems based on a splitting of
the spin Hamiltonian into two completely integrable components corresponding to ${\mathcal B}$-partitioned subsystems.
Further, we have implemented several variants of this integrator for a selection of small spin systems
and performed certain tests and comparisons. The results largely conform with the expectations; an
interesting finding is that, for fixed runtime, higher order algorithms yield marked improvements of the precision.
This accords with the results of \cite{HLW}, section V.3.2, where, however, no further
improvement occurs beyond the order of $8$. \\
Of course, these tests are only preliminary and should be extended
to include, for instance, more spin systems, the longtime behavior
and the influence of different decompositions of the Hamiltonian.
Also we have not compared our method with other methods which are
energy- and volume-preserving, but not symplectic
\cite{FHL,KBL,TKL}. Our method cannot be applied to an arbitrary
Hamiltonian spin system without taking additional measures. This
is a draw-back, but simultaneously an advantage since it means
that one has to adapt the method for a given system in order to
find an optimal algorithm. In view of the applicability the
perhaps most pressing generalization would be to consider the case
of more than two integrable components of the Hamiltonian.
\section*{Acknowledgement}
We thank P.~Hage, M.~Krech, and S.-H.~Tsai for interesting
discussions and useful remarks on an earlier version of the
manuscript and D.~P.~Landau for drawing our attention to some
relevant literature.




\begin{thebibliography}{99}

\bibitem{VH}L.~van Hove, Time-dependent correlations between spins and neutron scattering in ferromagnetic crystals,
    Phys.~Rev.~{95} (1954) 1374
\bibitem{HLW} E.~Hairer, C.~Lubich, G.~Wanner, { Geometric Numerical Integration}, Springer, New York, 2002
\bibitem{VER} L.~Verlet,  Computer ``experiments" on classical fluids, I. Thermodynamical properties
    of Lennard-Jones molecules, Phys.~Rev.~{ 159} (1967) 98
\bibitem{ABL} M.~J.~Ablowitz, J.~F.~Ladik, {A nonlinear difference scheme and inverse scattering},
    Studies in Appl.~Math.~{ 55} (1976) 213
\bibitem{IKS} A.~L.~Islas, D.~A.~Karpeev, C.~M.~Schober, Geometric
    integrators for the nonlinear Schr\"odinger equation,
    J.~Comput.~Phys.~173 (2001) 116
\bibitem{FHL} J.~Frank, W.~Huang, B.~Leimkuhler, Geometric integrators
    for classical spin systems, J.~Comp.~Phys.~{ 133} (1997) 160
\bibitem{OMF1} I.~P.~Omelyan, I.~M.~Mryglod, R.~Folk, Algorithm for
    molecular dynamics simulations of spin liquids,
    Phys.~Rev.~Lett.~{ 86 } 5 (2001) 898
\bibitem{OMF2} I.~P.~Omelyan, I.~M.~Mryglod, R.~Folk, Molecular dynamics simulations of spin and pure liquids with
    preservation of all the conservation laws, Phys.~Rev.~E~{ 64} 1 (2001)  016105
\bibitem{KBL} M.~Krech, A.~Bunker, D.~P.~Landau, Fast Spin Dynamics Algorithms for Classical Spin Systems,
      Comput.~Phys.~Commun. { 111} (1998) 1
\bibitem{TKL} S.~Tsai, M.~Krech, D.~P.~Landau, Symplectic integration methods in molecular and spin dynamics,
    Braz.~J~.Phys.~{ 40} 2 (2004) 384
\bibitem{AGK} M.~Ameduri, B.~Gerganov and R.~A.~Klemm, Classification of integrable clusters of
    classical Heisenberg spins, {\sl Preprint} cond-mat/0502323
\bibitem{SS}R.~Steinigeweg, H.-J.~Schmidt, Classes of integrable spin systems, {\sl Preprint} math-ph/0504009
\bibitem{ARN} V.~I.~Arnold, { Mathematical Methods of Classical Mechanics},
        Springer, New York, 1978
\bibitem{AM} R.~Abraham, J.~E.~Marsden, { Foundations of Mechanics}, 2nd edition, Addison-Wesley, London, 1978
\bibitem{AMR} R.~Abraham, J.~E.~Marsden, T.~S.~Ratiu, { Manifolds, Tensor Analysis, and Applications},
    Addison-Wesley, London, 1983
\bibitem{TR} H.~F.~Trotter, On the product of semi-groups of operators,\\ Proc.~Am.~Math.~Soc.~{ 10} (1959) 545
\bibitem{STR} G.~Strang, On the construction and comparison of difference schemes,
    SIAM J.~Numer.~Anal.~{ 5} (1968) 506
\bibitem{SU90}M.~Suzuki, Fractal decomposition of exponential operators with applications to many-body theories and
    Monte Carlo simulations, Phys.~Lett.~A{ 146} (1990) 319
\bibitem{FR} E.~Forest, R.~D.~Ruth, Fourth-order symplectic integration, Phys.~D { 43} (1990) 105
\bibitem{OMF} I.~P.~Omelyan, I.~M.~Mryglod, and R.~Folk, Comput.~Phys.~Commun.~{ 146} (2002) 188
\bibitem{WEB} Webpage http://mathworld.wolfram.com/topics/Polyhedra.html
\bibitem{FE30}A.~M\"uller, et al., Archimedean synthesis and magic numbers: ``Sizing" giant
    molybdenum-oxide-based molecular spheres of the Keplerate type, Angew.~Chem.~, Int.~ Ed.~{ 38} (1999) 3238

\end{thebibliography}
\end{document}